\documentclass{aastex}
\usepackage{emulateapj5}

\usepackage{epsfig}

\newcommand{\kms}{km\,s$^{-1}$}
\newcommand{\hi}{H\,{\sc i}}
\newcommand{\lya}{Lyman~$\alpha$}
\newcommand{\si}{S\,{\sc i}}
\newcommand{\siii}{Si\,{\sc ii}}
\newcommand{\siiii}{Si\,{\sc iii}}
\newcommand{\siiv}{Si\,{\sc iv}}
\newcommand{\ci}{C\,{\sc i}}
\newcommand{\cii}{C\,{\sc ii}}
\newcommand{\civ}{C\,{\sc iv}}
\newcommand{\oi}{O\,{\sc i}}

\newcommand{\nv}{N\,{\sc v}}

\begin{document}

\title{Detection of oxygen and carbon in the hydrodynamically escaping
atmosphere of the extrasolar planet HD\,209458b}

\author{A.~Vidal-Madjar\altaffilmark{1},
J.-M.~D\'esert\altaffilmark{1},
A.~Lecavelier des Etangs\altaffilmark{1},
G.~H\'ebrard\altaffilmark{1},
G.~E.~Ballester\altaffilmark{2},
D.~Ehrenreich\altaffilmark{1},
R.~Ferlet\altaffilmark{1},
J.~C.~McConnell\altaffilmark{3},
M.~Mayor\altaffilmark{4}, 
C.~D.~Parkinson\altaffilmark{5}
}

\altaffiltext{1}{Institut d'Astrophysique de Paris, CNRS, 98bis boulevard 
Arago, F-75014 Paris, France;
alfred@iap.fr, desert@iap.fr, lecaveli@iap.fr, hebrard@iap.fr, 
ehrenrei@iap.fr, ferlet@iap.fr}
\altaffiltext{2}{Lunar and Planetary Laboratory, Univ. of Arizona, 1040 E.
4$^{\rm th}$ St., Rm~901, Tucson, Arizona 85721-0077, USA;
gilda@vega.lpl.arizona.edu}
\altaffiltext{3}{Department of Earth and Atmospheric Science,
York University, North York, Ontario, Canada;
jack@nimbus.yorku.ca}
\altaffiltext{4}{Observatoire de Gen\`eve, CH-1290 Sauverny, Switzerland;
michel.mayor@obs.unige.ch}
\altaffiltext{5}{Jet Propulsion Laboratory/California Institute of
Technology and the NASA Astrobiology Institute, MS 150-21, 1200
E. California Blvd.  Pasadena, CA 91125, USA; cdp@gps.caltech.edu}

\begin{abstract}
Four transits of the planet orbiting the star HD\,209458 were observed
with the STIS spectrograph on board HST. The wavelength domain
(1180-1710\AA ) includes \hi\ as well as \ci , \cii , \civ , \nv , 
\oi , \si , \siii, \siiii\ and \siiv\ 
lines. During the transits, absorptions are
detected in \hi , \oi\ and \cii\ (5$\pm$2\%, 13$\pm$4.5\% and 7.5$\pm$3.5\%,
respectively). No absorptions are detected for other lines. The 5\% mean
absorption over the whole \hi\ \lya\ line is consistent with the previous
detection at higher resolution (Vidal-Madjar et al.\ 2003). The absorption
depths in \oi\ and \cii\ show that
oxygen and carbon are present in the extended upper atmosphere of
HD\,209458b. These species must be carried out up to the Roche lobe and
beyond, most likely in a state of hydrodynamic escape.
\end{abstract}


\section{Introduction}

The extrasolar planet 
HD\,209458b is the first one for which
repeated transits across the stellar disk have been observed ($\sim$1.5\%\ 
absorption; Henry et al.\ 2000; Charbonneau et al.\ 2000). Together with
radial velocity measurements (Mazeh et al.\ 2000), this has led to a
determination of the planet's radius and mass, confirming that it is a gas
giant. 
During transits, the
Hubble Space Telescope (HST) allowed the detection of the dense lower
atmosphere in the neutral sodium lines ($\sim$0.02\%\ additional
absorption, Charbonneau et al.\ 2002) and the extended upper atmosphere
in H\,{\sc i} ($\sim$15\% absorption over the stellar \lya\ emission 
line, Vidal-Madjar et al.\ 2003).
This \hi\ absorption extends
beyond the Roche lobe, showing a population of escaping atoms. 
The escape rate is
estimated to be $\ga 10^{10}$\,g\,s$^{-1}$, consistent with theoretical
evaluations (H\'ebrard et al.\ 2003, Lammer et al.\ 2003, Lecavelier des
Etangs et al.\ 2004). 
%
%
The additional HST observations which we present here
allow the upper atmosphere of HD\,209458b to be further studied.

\section{Observations and data reduction}
\label{observations}

Four transits of HD\,209458b were surveyed with the STIS G140L spectrograph 
with four HST visits on Oct.~9, Oct.~21, Nov.~5 and Nov.~24, 2003.
For each transit, three exposures during three consecutive HST orbits were
scheduled such that at least one full exposure was obtained with the
planet in front of the star. A previous or a following exposure, either
entirely before the first contact or after the last contact, serves as
a~reference.

The resulting 1D and 2D spectra, output from the STIS pipeline, are
similar to those described by Vidal-Madjar et al.\ (2003). In particular
they show detectable geocoronal emissions in \hi\ and \oi\ lines and a
non-zero background level due to the dark current. To correct for these
effects, we extracted the spectra from the 2D images of the 
detector following the procedure developed by Vidal-Madjar et al.
(2003) and D\'esert et al.\ (2003). In each column of the detector we
evaluate the background signal, including the geocorona, above and below
the stellar spectrum and interpolate its value to be subtracted from the
stellar spectrum.
%
The errors due to the background and the 
geocorona are included in the error estimates.

The wavelength domain of the 
G140L first order grating ranges
from 1180\,\AA\ to 1710\,\AA . 
The 
spectra are dominated by various stellar emission lines
(Si\,{\sc iii} $\lambda$1206\AA ,
\hi\ $\lambda$1216\AA ,
\nv\ $\lambda$1239\AA\ and $\lambda$1243\AA , 
\oi\ $\lambda$1302\AA ,
\cii\ $\lambda$1335\AA ,
Si\,{\sc iv} $\lambda$1394\AA\ and $\lambda$1403\AA ,
\si\ $\lambda$1474\AA ,
\siii\ $\lambda$1527\AA ,
C\,{\sc iv} $\lambda$1548\AA\ and $\lambda$1551\AA , C\,{\sc i}
$\lambda$1560\AA\ and $\lambda$1657\AA) 
and the stellar continuum is detected down to the shortest wavelength
(Fig.~1). The low resolution 
($\sim$2.5\,\AA )
does not allow the stellar emission lines to be resolved. The spectral
information is consequently limited to the total line intensity
variations.  However, this allows for high sensitivity and large
spectral coverage to search for many different species in the
planetary upper atmosphere.

\section{Analysis}
\label{analysis}

To visualize the spectral signature of the transit, we over-plotted
off-transit and in-transit spectra (Fig.~2). A significant difference is
seen in \hi , \oi\ and \cii , corresponding to an absorption during the
transit. Other spectral lines with about the same intensity do not show
the same behaviour (e.g., Si\,{\sc iii} in Fig.~2). This is the first
indication that, in addition to \hi , oxygen and carbon are also present
in the upper atmosphere of HD\,209458b.

To quantitavely estimate the 
absorptions, their level of detection
and error bars, we fitted transit profiles to the measured intensity
of each line 
in each exposure as a function of the orbital
phase.  The depth and width of the transit curve are related to
$R_{\rm abs}$/$R_{*}$, the ratio of the occulting object
and stellar radii. The absorption depth due to the planetary disk
is $(R_{\rm abs}$/$R_{*})^2\sim 1.5\%$. 
The duration of the ingress
and egress of the occultation 
(i.e. the width of the
transit curve) is taken to be proportional to $R_{\rm abs}$. 
The term 
$R_{\rm abs}$/$R_{*}$ is the first free parameter of the fit.

The intensity of the 
emission lines shows detectable variations
from one HST visit to another. These variations are interpreted as stellar
variations, known to occur in a solar type star at a few percent
level on a time scale of several~days. 
To extract the transit signature from
these intrinsic variations, we consider the lines intensity as a free
parameter for each of the 4 visits. We have thus 4 additional free
parameters for each spectral line. The intrinsic stellar variations are
found to be at a level of only a few percent between different~visits.

The stellar emissions 
during 3 consecutive exposures of~a given HST visit 
are assumed to be constant. We consider unlikely that 
short time-scale variations could mimic a
transit curve in \hi , \oi\ and \cii\ and not in the 
other lines, particularly those of the more ionized species 
for which more pronounced stellar variations should be observed.

We measured the total line intensities after subtraction of
the stellar continuum. 
The resulting 12~measurements have been fitted as a
function of the orbital phase by an occultation curve with five~free
parameters.
Results are shown in Fig.~3 and summarized in Table~1. The full continuum
spectrum from 1350\AA\ to 1700\AA\ shows a transit detection with a depth
of 2.0$^{+0.5}_{-0.7}$\%. In lines, only H\,{\sc i}, O\,{\sc i} and
C\,{\sc ii} show a detection at more than 2-$\sigma$ with transit depths
of 5$\pm$2\% in H\,{\sc i}, 13$\pm$4.5\% in O\,{\sc i} and 7.5$\pm$3.5\% in
C\,{\sc ii} (2.6, 2.6 and 2.1-$\sigma$ detection, respectively). 
The total squared difference between the line intensities and the fits
are found to have values of 8.9, 9.6 and 4.7 (for \hi , \oi\ and 
\cii , respectively), in agreement with
a $\chi^2$ distribution with 7 degrees of freedom.
The other stellar features do not show detectable signatures during
transit. 

A 10\% depth corresponds to a radius $R_{\rm abs}$ of about
3.6~Jupiter radii, which is also the radius of the Roche lobe. The
measured absorption depths in \oi\ and \cii\ correspond to absorbing 
clouds with a size of several planetary radii. The present detections 
show that oxygen and carbon are present in the extended upper atmosphere; 
\oi\ is even present up to about the level of the Roche lobe.

\section{Discussion: observational results}

\subsection{Continuum}

The detection of an absorption of 2.0$^{+0.5}_{-0.7}$\% over the continuum
validates the adopted procedure. This absorption is due to the transit of
the planetary disk and is known to be $\sim$1.5\%, well within our error bar.
This shows that the systematic errors are significantly below this level.

\subsection{Atomic hydrogen}
\label{hydrogen}

Vidal-Madjar et al.\ (2003) detected an \hi\ absorption depth of
15$\pm$4\% over the Lyman-$\alpha$ line. The absorption is significantly
detected over only part of the Lyman-$\alpha$ line within a radial
velocity range from $-130$ to $100$\,\kms (Fig.~2 of Vidal-Madjar et al.
2003). The 15$\pm$4\% absorption in that spectral range corresponds to
about 5.7$\pm$1.5\% absorption of the total Lyman-$\alpha$ line intensity. 
The \hi\ absorption depth measured here is consistent with the results of
Vidal-Madjar et al.\ (2003). 

\subsection{Oxygen and carbon}
\label{oi and cii}

The \oi\ stellar emission triplet is a blend of the ground level line at
1302.2\,\AA\ with the lines from excited levels O\,{\sc i}* and
O\,{\sc i}** ($\lambda$=1304.9\,\AA, energy level $E$=158\,cm$^{-1}$ and
$\lambda$=1306.0\,\AA, $E$=227\,cm$^{-1}$, respectively). The \oi\ 
ground level line is strongly absorbed by the interstellar medium (Fig.~2).
Therefore the $\sim$13\% 
absorption observed during the transit in the
full \oi\ triplet must be due to the presence of \oi* and \oi** within
several planetary radii around the planet, up to about the radius of the
Roche lobe. Because these excited levels of \oi\ are likely populated by
collisions, the \hi\ 
density at this altitude should be at least the density of a typical
exobase, that is about 10$^6$\,cm$^{-3}$.  Assuming solar abundances,
such a density is needed for the \oi\ triplet to have enough opacity
to be detected in a line of sight grazing the planet.  In these
conditions, only the strongest lines of the most abundant species
in the most abundant ionization state can be detected, 
i.e. \hi, \oi\ and \cii\ (C\,{\sc i} is ionized above~11.3\,eV). 

Seen at low resolution, the stellar \cii\ line is also a blend of a
ground level line from C\,{\sc ii} ($\lambda$=1334.5\,\AA) with a line from
C\,{\sc ii}* ($\lambda$=1335.7\,\AA, $E$=63.42\,cm$^{-1}$). Since the
ground level C\,{\sc ii} line is only partially absorbed by the interstellar
medium (Fig.~2), at the present low spectral resolution, we cannot a
priori discriminate between absorption by \cii\ or \cii *. However, for
densities above $\sim 20$\,cm$^{-3}$ (Wood \& Linsky 1997), \cii * is
expected to be collisionally populated, we thus conclude that the detected
absorption is due to the presence of both \cii\ and \cii * in the extended
upper atmosphere of HD\,209458b.

Due to the lack of spectral resolution, the real absorption depth over 
a wavelength limited domain of an emission line (which is 
linked to the size of the occulting 
cloud) can be different from the measured 
absorption depth over the full stellar emission intensity 
(Sect.~\ref{hydrogen}). The measured decrease in intensity
depends on the velocity dispersion of the absorbing species compared to
the intrinsic width of the stellar emission 
lines. HD\,209458 being a solar type star, we
can use the Sun as observed by the SUMER
instrument as a reference; the width of the \oi\ and \cii\ stellar lines
must be about 
$\sim$15\,km\,s$^{-1}$ for \oi\ and
$\sim$25\,km\,s$^{-1}$ for \cii\ (Chae et al.\ 1998; P. Lemaire, private
communication). These values are consistent with the high resolution
echelle spectra of HD\,209458 (Fig.~2). Therefore, in the absorbing cloud, 
velocity dispersions
of, for example, 25, 15 and 3\,km\,s$^{-1}$ lead to corrected absorption
depths of 13$\pm4.5$\%, 13$\pm4.5$\% and 65$\pm23$\% for \oi\ and
7.5$\pm3.5$\%, 12.5$\pm6$\% and 62.5$\pm30$\% for \cii , respectively. 
Note that we obtain similar absorption depths for \oi\ and \cii\ 
for velocity dispersions below $\sim$15\,km\,s$^{-1}$. 

\section{The escape mechanism}

As shown by Vidal-Madjar et al (2003), hydrogen atoms are escaping away
from HD\,209458b. This has been further discussed by H\'ebrard et al.\ 
(2003), Lammer et al.\ (2003), Liang et al.\ (2003; 2004), Lecavelier 
des Etangs et al.\ (2004) and Yelle (2004). 
The present detection of oxygen and carbon at high
altitudes provides additional information about the type of escape 
mechanism involved. 
First, oxygen is present up to about the Roche lobe where the \hi\ density
must be at least 10$^6$\,cm$^{-3}$. Second, the absorption depths of oxygen
and carbon are related to their velocity dispersion, with $\sim$13\%\
depth for 15\,km\,s$^{-1}$ velocity dispersion and $\sim$65\%\ depth for
3\,km\,s$^{-1}$.

\subsection{Jeans escape}

In the frame of Jeans escape, atoms and molecules diffuse upward from the
base of the thermosphere up to the exobase level above which the absence
of collision allows particles with large velocities in the tail of the
Maxwellian distribution to escape. 
In the thermosphere, the scale height of \hi\ is of the order of 10\,000~km 
(0.1~planetary radius).
Under molecular diffusion conditions, the \oi\ scale height is 16~times
smaller than the \hi\ one.
For an altitude increase of one \hi\ scale height, the O/H ratio decreases
by a factor e$^{16}/$e. Therefore, if molecular diffusion dominates no OI
can be detected at high altitudes. 

However, in the solar system planetary atmospheres, eddy diffusion is able 
to transport upward species within the thermosphere up to some level where
molecular diffusion takes over. This transport mixes the species, and 
their abundances remain unchanged upward. Assuming an extreme case of eddy 
transport up to the Roche lobe or to the
exobase, oxygen and carbon abundances should be solar at that level.
However even with eddy diffusion and upper
atmospheric temperature $T_{\rm up}\sim$10\,000\,K (Lecavelier des Etangs
et al.\ 2004), thermal broadening of absorption lines are
$\sim$13\,km\,s$^{-1}$ for \hi , and only $\sim$3\,km\,s$^{-1}$ for \oi\ 
and \cii . This is smaller by a factor of $\ga$5 than the width of the
stellar lines. If the absorption width was limited by the thermal
broadening, the observed $\sim$10\% 
absorption would correspond to a geometrical
absorption of more 60\%. This would imply that oxygen and carbon 
extend much further than \hi , which is unlikely. Eddy diffusion transport 
cannot explain the
properties of the 
absorption in \oi\ and~\cii .

\subsection{Hydrodynamic escape: ``blow-off''}

To explain the present observation, we need a mechanism able to transport
oxygen and carbon up to about the Roche lobe and to maintain their
velocity dispersion to at least $\sim$10\,km\,s$^{-1}$.
When an abundant species like \hi\ escapes
with a high rate, other species can be dragged up by the hydrodynamical 
flow, also called ``blow-off'' (see, e.g., Chamberlain \& Hunten 1987). 
Under such conditions,
oxygen and carbon moving out with the \hi\ flow should present a 
velocity dispersion at least similar to or greater than the sound speed, 
which is $\sim$10\,km\,s$^{-1}$ at 10\,000\,K, i.e. transsonic 
(Parkinson et al.\ 2004). The abundances of oxygen and carbon
should also be preserved upward all along the flow. Under these
conditions, the amounts of \oi\ and \cii\ are large enough to absorb the
whole stellar line, allowing at least 10\%\ absorption if they flow up to
the Roche lobe. Although the \oi\ and \cii\ absorptions are compatible
within the error bars, the dynamical ``blow-off'' could also explain the
apparent lower absorption in carbon as the \cii\ stellar emission
line is broader than the \hi\ sound speed by a factor of $\sim$2. In
short, the depth of the \oi\ and \cii\ absorptions can be interpreted
through the absorption by oxygen and carbon flowing away in a
hydrodynamic escape of the atmospheric hydrogen.

\section{Conclusion}

The still unique case of the planet HD\,209458b transiting a bright
star has allowed the first studies of an extra-solar planet
atmosphere. The detection of oxygen (at~2.6\,$\sigma$) and carbon (at
2.1\,$\sigma$), a signature of an atmospheric ``blow-off'', shows that,
as already suggested by H\'ebrard et al.\  (2003) and 
Vidal-Madjar \& Lecavelier des Etangs (2003) 
hot-Jupiters can lose a significant fraction of their
atmosphere. Ultimately, such a process could lead to a new type of
planets with hydrogen--poor atmosphere, or even with no atmosphere at
all (see Trilling et al.\ 1998).

\acknowledgments 
We thank the Egyptian god Osiris for suggesting his name for the planet
(Vidal-Madjar \& Lecavelier des Etangs 2003).
Based on observations obtained with the NASA/ESA 
Hubble Space Telescope operated by the Association of Universities
for Research in Astronomy, Inc., under NASA contract NAS 5-26555. We
thank S.~Beckwith
for allocating us Director
Discretionary time.  C.\,D.\,P.  acknowledges 
support
by the NASA through the Astrobiology
Institute under Cooperative Agreement No. CAN-00-OSS-01 and issued
through the Office of Space Science.

%
%
%
%
%
%

\clearpage

\begin{table}
\caption{Absorption depth of emission lines and continuum}
\hspace{4.5cm}
\begin{tabular}{lcrr}
\hline
\hline
Species & $\lambda$$^{\rm a}$ & 
\multicolumn{2}{c}{Absorption depth$^{\rm b}$} \\
 & (\AA ) & 1$\sigma$ (\%) & 2$\sigma$ (\%) \\
 \hline
Continuum    & 1350-1700 & $2.0^{+0.5}_{-0.7}$ \\
\\
H\,{\sc i}    & 1212-1220 &  $5.3^{+1.6}_{-1.9}$ \\
O\,{\sc i}, O\,{\sc i}*, O\,{\sc i}**& 1300-1310 &  $12.8^{+4.5}_{-4.5}$ \\
C\,{\sc ii}, C\,{\sc ii}*   & 1332-1340 & $7.5^{+3.6}_{-3.4}$ \\
\\
C\,{\sc i}    & 1557-1565 & $0.4^{+21.1}_{-0.4}$  & $<$36.0 \\ 
C\,{\sc i}    & 1654-1660 & $11.3^{+11.9}_{-11.3}$  & $<$33.6 \\ 
C\,{\sc iv}   & 1545-1554 & $0.4^{+9.5}_{-0.4}$  & $<$19.0 \\ 
N\,{\sc v}    & 1237-1246 & $27.3^{+22.7}_{-27.3}$  & $<$50.0\\ %
S\,{\sc i}, S\,{\sc i}*, S\,{\sc i}**& 1471-1489 & $32.2^{+13.0}_{-32.2}$  
& $<$58.2\\ 
Si\,{\sc ii}, Si\,{\sc ii}*& 1525-1536 & $18.4^{+15.4}_{-18.4}$  & $<$47.4\\ 
Si\,{\sc iii} & 1204-1210 & $0.0^{+2.2}_{-0.0}$  & $<$5.9 \\ 
Si\,{\sc iv}  & 1391-1397 & $0.0^{+6.5}_{-0.0}$  & $<$14.0 \\ 
\hline
\end{tabular}\\

\hspace{2.5cm}
$^{\rm a}$ Wavelength range for the lines intensity evaluation.

\hspace{2.5cm}
$^{\rm b}$ Absorption depth given by the $(R_{\rm abs}/R_*)^2$ parameter 
of the fit (see text). 

\hspace{2.5cm}
1-$\sigma$ error bars and 2-$\sigma$ upper limits for non-detections.

\end{table}

\clearpage

\begin{figure}
\hspace{4cm}
\psfig{file=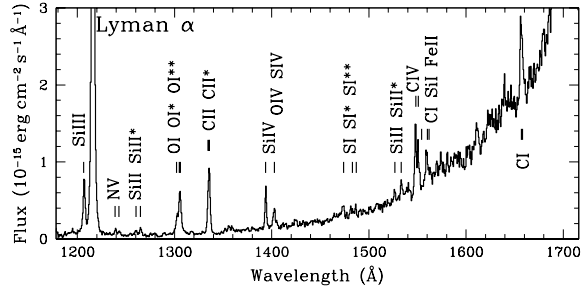,width=3.in}
\caption{The full HD\,209458 G140L spectrum. Stellar emission lines
are clearly detected together with the stellar continuum which is 
increasing towards longer wavelengths.
\label{full spectrum}}
\end{figure}

\begin{figure}
\hspace{4.5cm}
\psfig{file=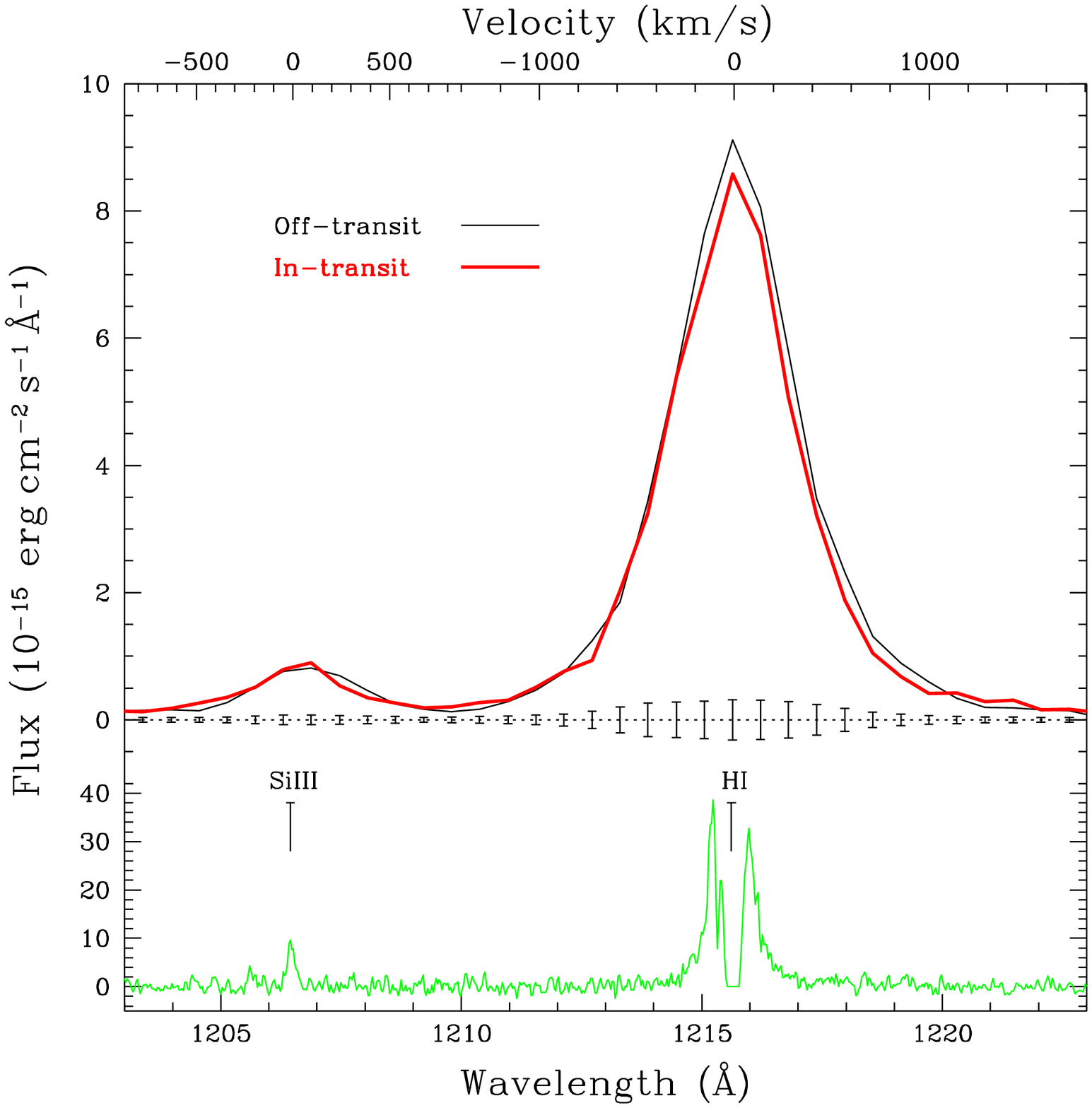,width=3.in}\\

\hspace{4.5cm}
\psfig{file=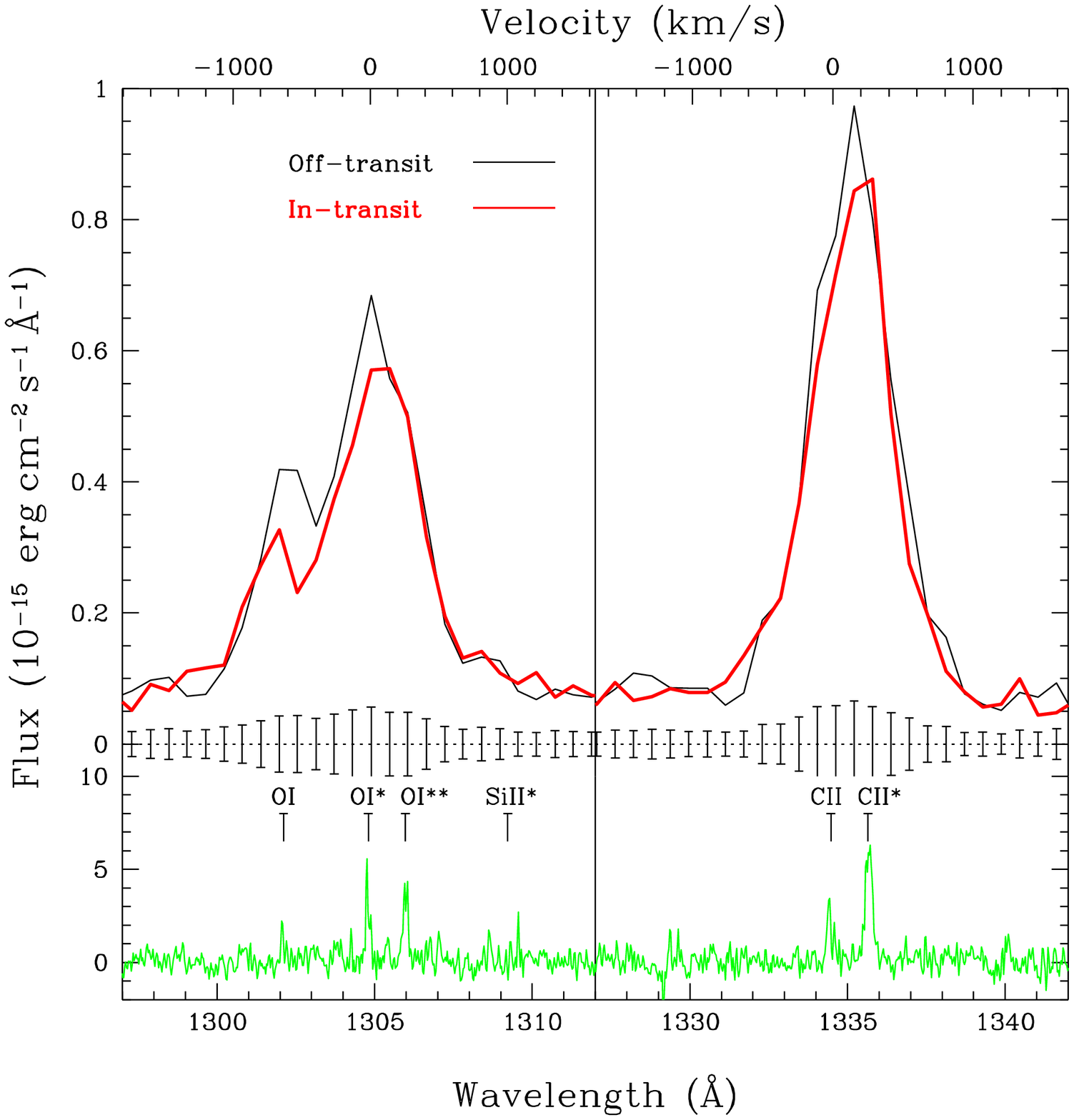,width=3.in} 
\caption{Comparison of off-transit
and in-transit spectra for the \siiii , \hi , \oi\ and \cii\ lines 
with 1-$\sigma$ error bars. The
off-transit spectrum is the addition of the first exposure of the 1$^{\rm
st}$ and 3$^{\rm rd}$ visits (thin lines); the in-transit corresponds to
the addition of the fully in-transit exposure of each of the four HST
visits (thick lines).
Absorption is clearly detected in the \hi , \oi\ and \cii\ lines. No
signal is detected in other lines (e.g., \siiii ) with about the same
amplitude. The bottom panels show the high resolution spectra obtained in
2001 with the echelle grating (Vidal-Madjar et al.\ 2003).}
\end{figure}

\begin{figure}
\hspace{4.4cm}
\psfig{file=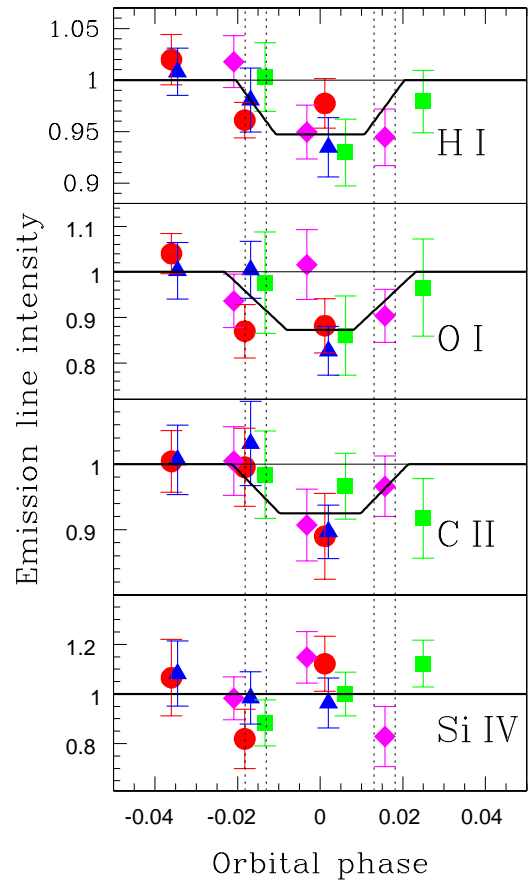,width=2.7in} 
\caption{Plot of the lines
total intensity as a function of the orbital phase. Circles, squares,
triangles and diamonds are for 1$^{\rm st}$ to 4$^{\rm th}$ transits
respectively. The vertical dotted lines corresponds to the position of the
1$^{\rm st}$ to 4$^{\rm th}$ contacts of the planetary disk transit. 
The thick line represents the best fit to the data 
(see Sect.~\ref{analysis}). Absorptions are detected in \hi , \oi\ and 
\cii\ during the transits; no significant absorptions are detected in the 
other lines (i.e. Si\,{\sc iv}).
\label{plot fit}}
\end{figure}

\end{document}